\documentclass[useAMS,usenatbib]{mn2e}

\title[The myth of the molecular ring]
{The myth of the molecular ring}
\author[C. L. Dobbs]
{C. L. Dobbs\thanks{E-mail:
dobbs@astro.ex.ac.uk}$^{1,2,3}$ and
A. Burkert$^{1,2}$\footnote{Max-Planck fellow}\\
$^1$ Max-Planck-Institut f\"ur extraterrestrische Physik, Giessenbachstra\ss{}e, D-85748 Garching, Germany \\
$^2$ Universitats-Sternwarte M\"unchen, Scheinerstra\ss{}e 1, D-81679
M\"unchen, Germany \\
$^3$ School of Physics, University of Exeter, Stocker Road, Exeter EX4 4QL}
\usepackage{amssymb}
\usepackage{amsmath}
\usepackage{graphicx}
\usepackage{epsfig}
\usepackage{multirow}

\begin{document}
\date{\today}

\pagerange{\pageref{firstpage}--\pageref{lastpage}} \pubyear{0000}

\maketitle

\label{firstpage}
\begin{abstract}
We investigate the structure of the Milky Way by determining how
features in a spatial map correspond to CO
features in a velocity map. We examine structures including
logarithmic spiral arms, a ring 
and a bar.  We explore the available parameter space, including the pitch angle of
the spiral arms, radius of a ring, and rotation curve. 
We show that surprisingly, a spiral arm provides a better fit to the observed molecular ring
than a true ring feature. This is because both a
spiral arm, and the observed feature known as the molecular ring, are curved in velocity
longitude space. 
We find that much of the CO emission in the velocity
longitude map can be fitted by a nearly symmetric 2 armed spiral
pattern. One of the arms
corresponds to the molecular ring, whilst the opposite arm naturally reproduces the Perseus arm.
Multiple arms also contribute to further emission in the vicinity of the
molecular ring and match other observed spiral arms. Whether the Galactic structure consists primarily of two, or several spiral arms, 
the presence of 2 symmetric
logarithmic spirals, which begin in the vicinity of the ends of the
bar, suggest a
spiral density wave associated with the bar.
\end{abstract}

\section{Introduction}
Despite decades of observations, determining the spiral structure of
our Galaxy is still intrinsically difficult. It is not even clear
whether the Galaxy contains 2, 3 or 4 primary spiral arms
\citep{Vallee2005,Vallee2008,Benjamin2008,Steiman2010}, and whether the spiral structure
is different in the gas and the stars. Traditionally, much of the CO
emission of the Galaxy has been associated with a feature known as the `molecular
ring' \citep{Stecker1975,Cohen1977,Roman2010}, around 4 kpc from the
Galactic Centre. However it is unclear whether this is truly a ring, or simply
emission from nearby spiral arms, as suggested by simulations of
spiral galaxies \citep{Englmaier1999,Rod2008,Baba2010}.

In the past, the spiral structure for the Galaxy has predominantly
been determined from the stellar distribution (\citealt{Vallee2005} and
references therein). Measuring distances to stars is difficult for
distances larger than a few kpc however due to
extinction. Alternatively we can use a gas tracer such as CO or H{\sc{i}}. In other
galaxies, e.g. M83, M51, the spiral arms also tend to be much narrower
in the gas than the stars, indicating that gas is likely a better
tracer of spiral structure. For a spiral density wave,
 the gaseous and stellar spiral arms are
expected to occupy slightly different patterns, with the gaseous arms
slightly offset from the stellar arms except at corotation, and with a
smaller pitch angle \citep{Gittins2004} \footnote{although recent simulations
of M51 \citep{Dobbs2010} find that for a kinematic wave driven by a tidal interaction, the
stellar and gaseous arms are not systematically offset.}. The gaseous
structure is also much more complex than the stellar, with interarm
spurs, and branches between spiral arms absent in the stellar distribution.

CO and H{\sc{i}} maps of the Galaxy
(e.g. \citealt{Dame2001}) clearly show the Perseus and
Outer arms.
However there are also difficulties with using gas tracers: i) 
 it is difficult to map the opposite side of the Galaxy, ii) the emission in the
 inner part of the Galaxy is dominated by a broad band in
 velocity-longitude (hereafter $l-v$) space, and
 iii) ambiguities in calculating the distance to gaseous
   features from the rotation curve and velocity crowding. Thus mapping the
spatial structure from the gas is far from straightforward.

In this paper we take a slightly different approach. Rather than using
the molecular emission, or stellar distribution, to estimate the
spiral structure, we instead assume the gaseous spiral arms exhibit some pattern and
see how well they fit the observed CO emission. We do not perform
numerical simulations rather we simply assume the gaseous spiral arms
follow a logarithmic spiral pattern, assumed to arise from
the gas response to a density wave. This has the caveat that our results neglect
streaming motions. However if we can fit the spiral pattern even in the
absence of streaming motions, this is a strong indication
that the spiral pattern for our Galaxy can be represented by a simple
$m=2$ or $m=4$ pattern. There are also two direct advantages of our method; the
first that we can readily explore a large parameter space, and the
second that we do not need to include the pattern speed, or spiral potential strength,
which are unknown parameters.
A similar approach has been
carried out by \citet{Russeil2003} for star forming complexes, but the
distribution they use does not display strong spiral
structure. \citet{Steiman2010} also fit spiral patterns to the 
intensity of FIR cooling lines at each position in the Galaxy.

\section{Method}
To obtain spiral arms, we assume that the molecular gas lies in a 2 or 4 armed logarithmic spiral
pattern. From standard density wave theory, the general expression for
a logarithmic spiral pattern is
\begin{equation}
\phi=A(r) \cos \bigg(\frac{n}{\tan i} \log(r/r_0)-(\theta-\Omega_p t)\bigg)
\end{equation} 
where $A(r)$ provides the amplitude of the spiral, $n$ is the number
of spiral arms, $i$ is the pitch angle, $r_0$ is a constant
  which controls the orientation of the arms, and
$\Omega_p$ the pattern speed. Since we only require the pattern at the
present time, we can set $t=0$.  To find the minima of the potential,
we then simply have 
\begin{equation}
\frac{n}{\tan i} \log(r/r_0)=\theta,
\end{equation}
and given some values of $i$ and $r_0$ we can map the positions of the
spiral arms.

We then compute the velocity longitude map, requiring a given rotation
curve. We adopt a flat rotation curve
for a logarithmic potential \citep{Binney1987}, of the form
\begin{equation}
v=v_0 r^2/(R_c^2+r^2)
\end{equation}
where $v_0$ and $R_c$ are constants. $R_c$ determines how far from
the centre of the Galaxy the rotation curve becomes flat.
Past observations have indicated the rotation of the Galaxy is between 210 and
240 km s$^{-1}$ over the majority of the Galaxy \citep{Clemens1985}.
The standard reference is $\Theta_0=220$ km s$^{-1}$ at $R_0$=8 kpc.
However more recent measurements of the distance to masers suggest
that $\Theta_0$ may be 250~km~s$^{-1}$ or higher \citep{Reid2009}. 
We tried both $v_0=$220 and 250~km~s$^{-1}$, though we only show
results for $v_0$=250 km s$^{-1}$.
 As the observations show that the
velocity curve is still very high close to the centre of the Galaxy,
we choose $R_c=0.1$ kpc. Then we
place the observer a distance of 8 kpc from the centre of the Galaxy.

We show results where we adopt a spiral arm pattern, and where
  we assume that the molecular ring is truly due to a ring.
Given that we simply use the computed locations of the spiral arms, or
ring, we
can investigate a large parameter space. 
The free parameters in our models are the pitch angle of the spiral
arms; the radius of a ring feature; the orientation of the spiral arms, the length and orientation
of the bar; the Galocentric radius and the rotation velocity $v_0$.
We mainly consider the orientation and pitch angle of the spiral arms,
but we also briefly mention the other parameters. In principle, this analysis could be carried out
without any prior knowledge of the structure of the Milky Way, but
given the large parameter space we have started with the location of
the observer, the rotation curve, and the orientation of the bar
roughly based on observations.

\subsection{Fitting technique}
We can compare how well our models match the CO observations
by matching features such as the molecular ring, Perseus Arm, Outer
Arm simply by eye. However we also carry out a $\chi^2$ fitting between the
velocity longitude map of \citet{Dame2001} and our models. The
difficulty of the latter is that we have to make numerous assumptions
to convert our models into emission maps.  
We assume the emission follows a Gaussian
centred on the proposed spiral arms with a velocity dispersion of 7 km
s$^{-1}$. We also have to make some assumptions about how the emission
scales with radius. We suppose the intensity falls with $1/r_{LSR}^2$
  where $r_{LSR}$ is the distance to the local standard of rest
  (located at $R=8$ kpc). We also assume that the amount of molecular
  gas falls off as $1/r_{GAL}$ where $r_{GAL}$ is the radius of the
  Galaxy. We then normalise the emission so that the total emission
  matches that of \citet{Dame2001}. We calculate the difference
  between the observed and model emission,
  $\sigma^2=(I_{obs}-I_{mod})^2$, and minimise over the spiral arm
  orientation or molecular ring radius. Whilst departures from these
  assumptions (e.g. a different velocity dispersion, changes in
  scaling with $r_{LSR}$ and $r_{GAL}$) change $\sigma$,
  how well the models fit relative to each other does not change.

There are still some difficulties with our fitting process when we
compare to spiral features. In the
Galaxy, local emission is present at all longitudes, but absent in
our models. Away from the Galactic Centre, this emission dominates
over features such as the Outer Arm. Thus we cannot really test for
these features without introducing some arbitrary weighting, so we
instead restrict our fit to
longitudes between 50$^{\circ}$ and -50$^{\circ}$. We still found however that
this method was biased towards lower pitch angles, simply because the 
arms cross the region $l=\pm50^{\circ}$ multiple times. Therefore we also carried
out a fit just to the part of the main spiral arm which coincides with the
molecular ring. We refer to the two fits as `total' (i.e.\ for all the
parts of the arms in our models between $l=$50$^{\circ}$ and -50$^{\circ}$) and
`arm' (i.e.\ just between the tangent points of the arm which
coincides with the molecular ring). 

In the first part of the results we compare a 2 armed and 4
armed spiral pattern, so we use the `total' fit. In Section 3.1 we
vary the pitch angle and compare results with both the `total' and
`arm' fits. The best fit orientation of the arm does not depend on
which technique is used, but the pitch angle does.

\section{Results}
In Figure~1 we show our best fit to the molecular ring for a
  2 armed spiral pattern (top panel) adopting a pitch angle of 11$^{\circ}$
  (see next section for results with different pitch angles). 
This model includes a
bar of radius 3 kpc, which we have simply placed across the Galactic
centre 45$^{\circ}$ clockwise from the position of the Sun. In all
  figures, we simply show the lines tracing peak emission along the
  arms, rather than our synthetic emission maps. The lines are 
 overplotted on the
  velocity longitude map of \citet{Rod2008}, which used the data of 
\citet{Dame2001}.
\begin{figure*}
\centerline{\includegraphics[scale=0.5, bb=-50 -100 400 360]{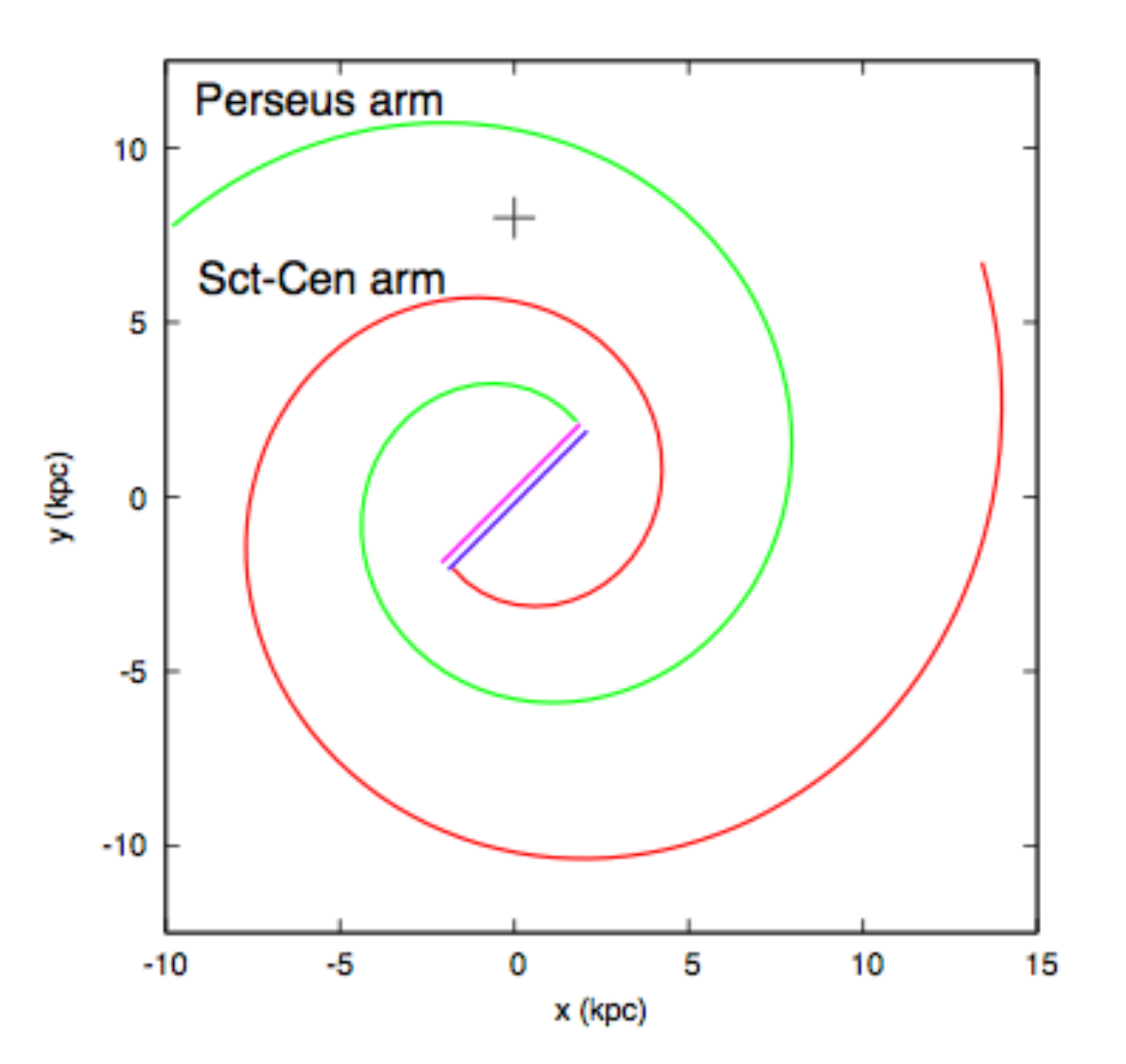}
\includegraphics[scale=0.58, bb=0 -100 530 310]{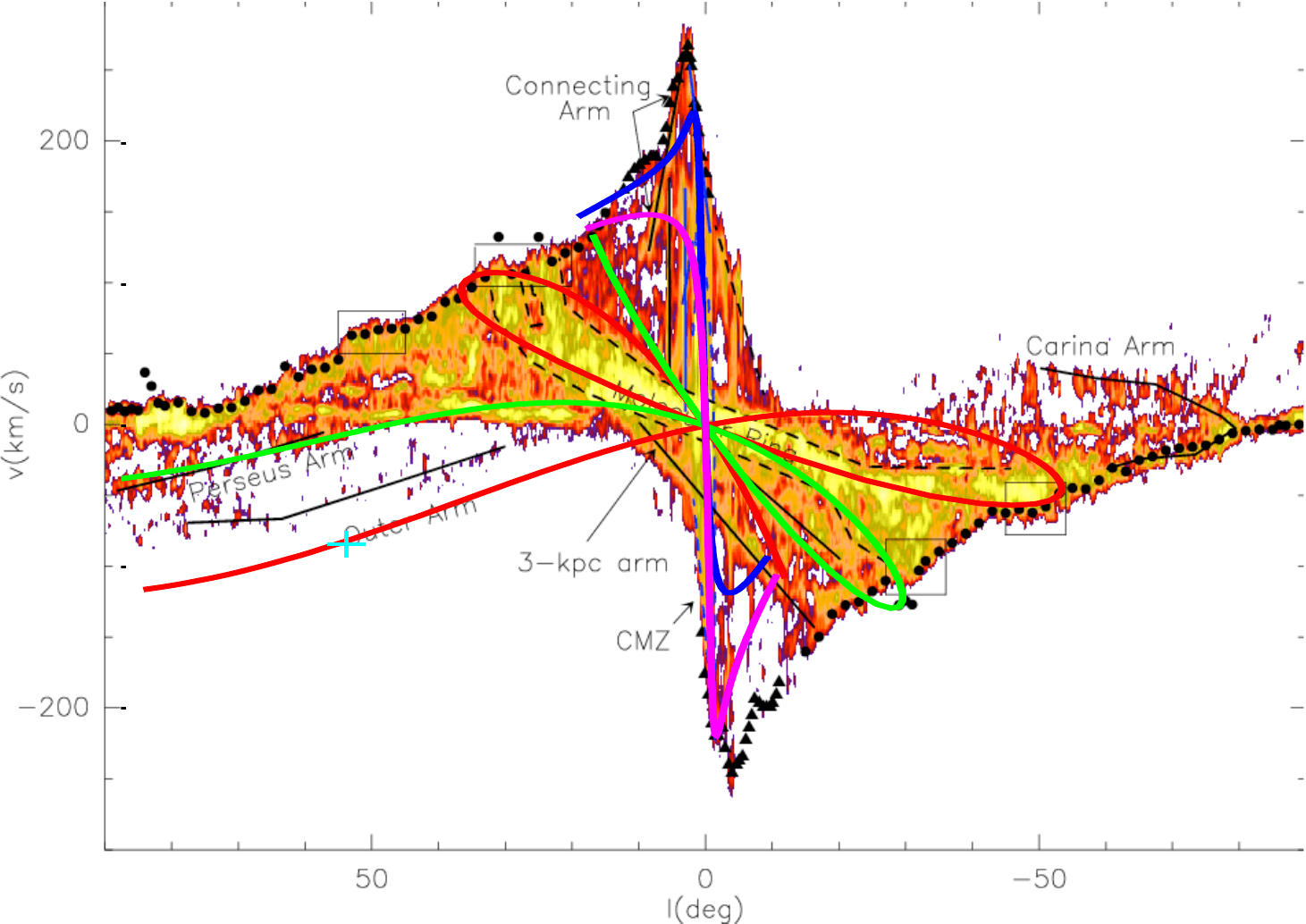}}
\centerline{\includegraphics[scale=0.5, bb=-50 -100 330 210]{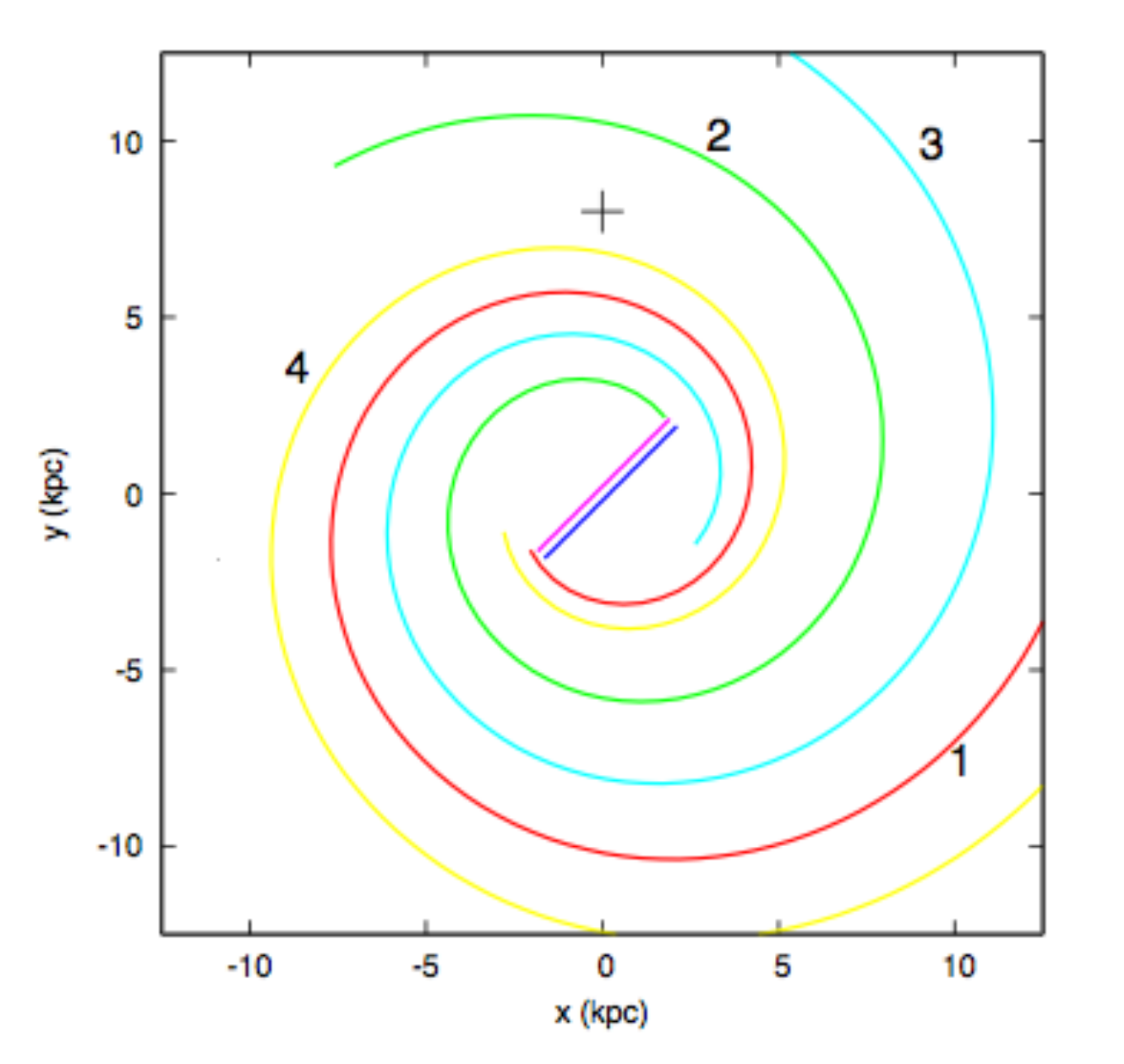}
\includegraphics[scale=0.58, bb=-60 -100 530 230]{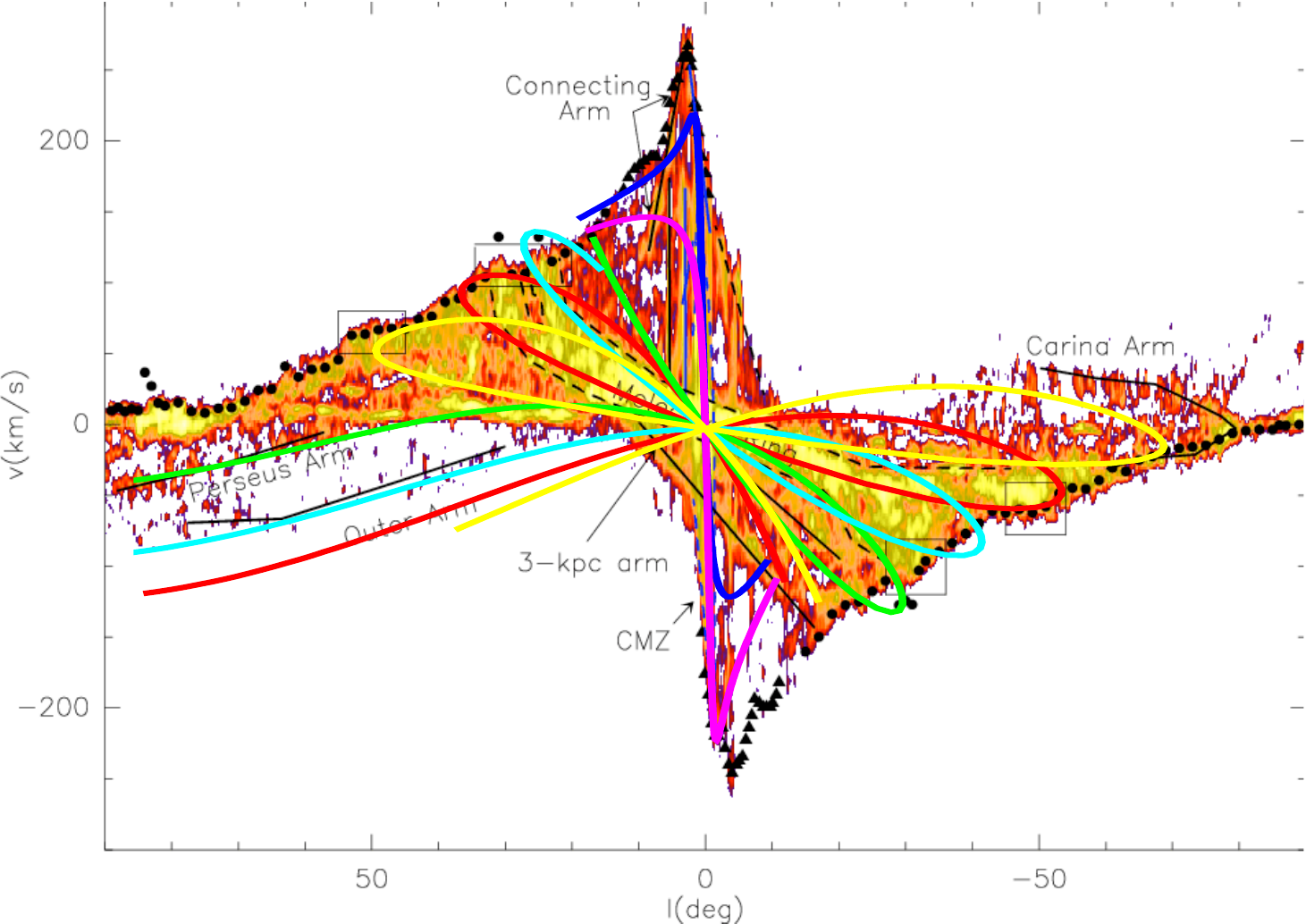}}
\centerline{\includegraphics[scale=0.39, bb=65 0 330 280]{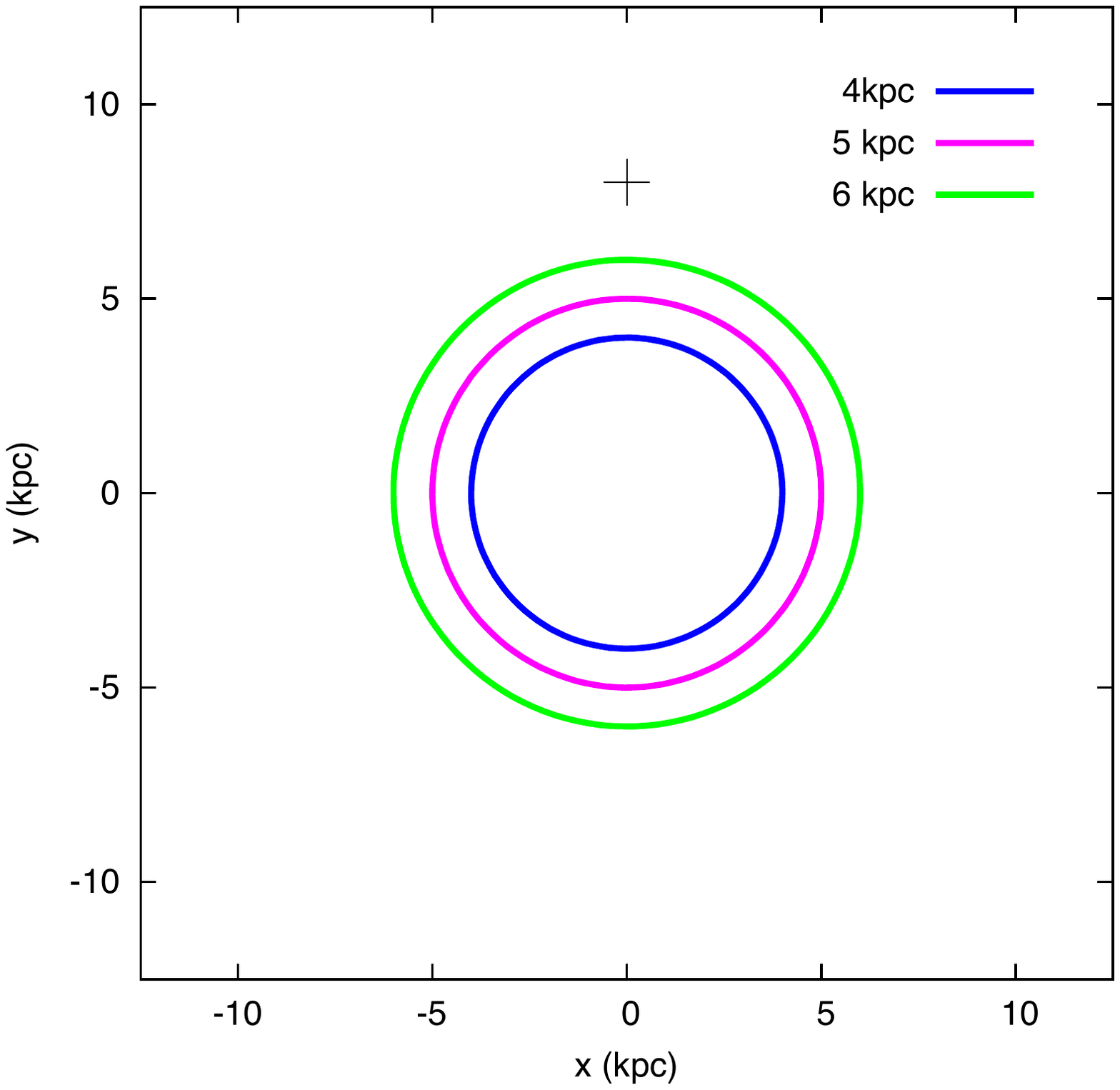}
\includegraphics[scale=0.58, bb=-200 -40 530 230]{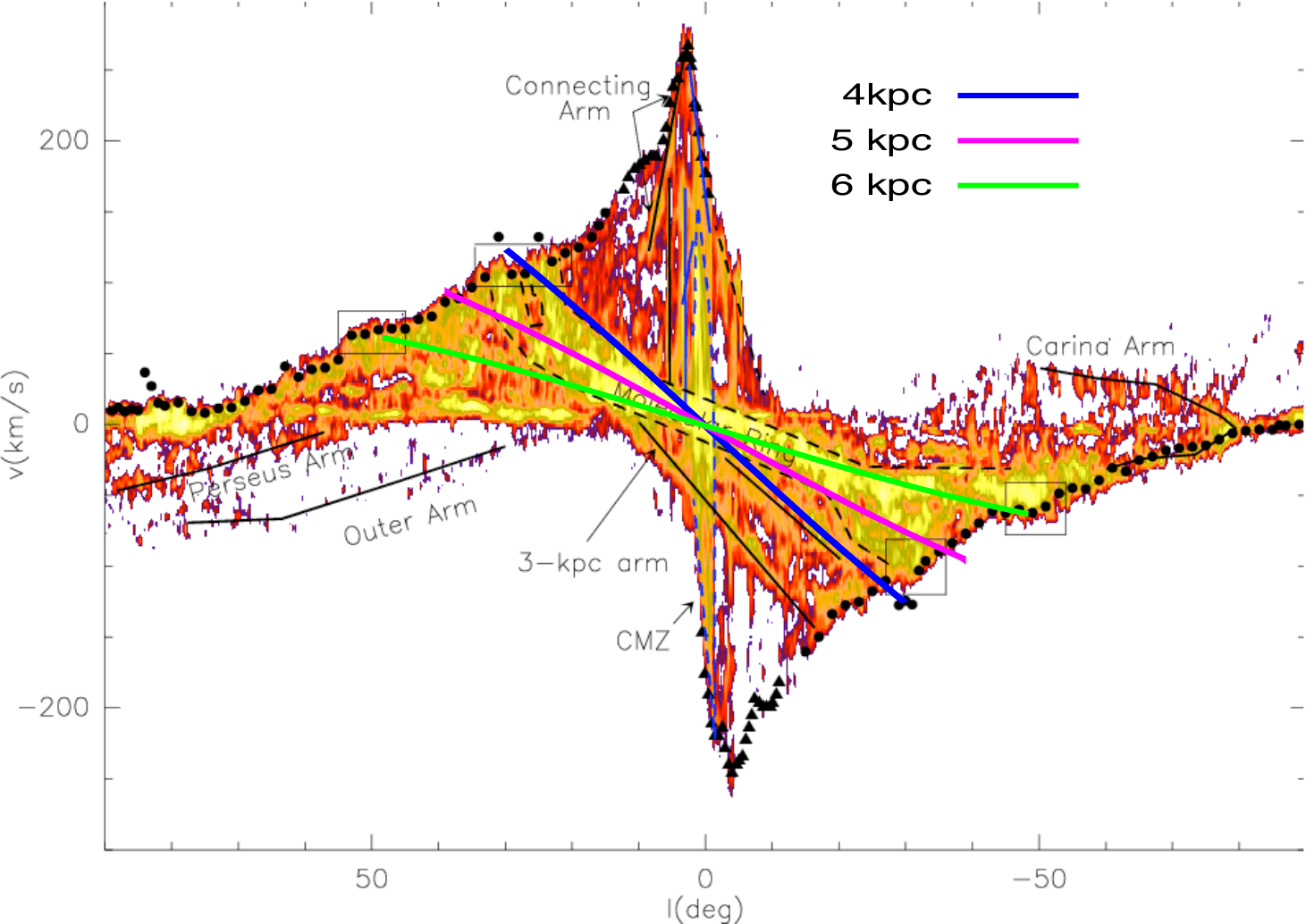}}
\caption{On the left hand side we illustrate possible Galactic
  features, a 2 armed spiral pattern (top), a 4 armed spiral pattern
  (centre) and 3 rings of different radii (lower). The position of the observer is marked by the cross at 8 kpc. On the right hand
  panels we show the location of the spiral arms in a velocity
  longitude plot, from \citet{Rod2008} who used the data of
  \citet{Dame2001}. The black circles and crosses indicate terminal
  velocity measurements. The cyan cross shown in the top right panel
  indicates the extremity of the outer HI arm observed by
  \citet{Dame2011}. The pitch angle for the spiral models is
  11$^{\circ}$. The spiral patterns are the best fit patterns (with regards
  to the orientation of the arms) to the observed emission between 
$l=\pm 50^{\circ}$. The Scutum-Centaurus arm provides a good fit to the
  molecular ring. The best fit ring has radius 6 kpc. However since
  the spiral arm is curved in $v-l$ space, 
it has a better fit to the molecular ring compared to a ring. Please view journal online for colour versions of the figures.}
\end{figure*}

From Figure~1, top panel, we see that the Scutum-Centaurus (hereafter
Sct-Cen) 
arm provides
  a good fit to the molecular ring. The part of this arm in the lower
  left quadrant (negative $l$ and $v$) also agrees very well with the
  location of the new arm as measured in HI by \citet{Dame2011}. As
  also noted in their paper, it is very difficult to get the Sct-Cen arm
  to continue to the Outer arm, without providing a very asymmetric
  spiral pattern. The second spiral arm provides a good fit to
  the Perseus arm. The best fit for the second arm was almost
  symmetric with the Sct-Cen arm, only asymmetric by 10$^{\circ}$. 
  It is possible to rotate the arms
  slightly and produce better matches to the tangent points (boxes), still
  matching the spiral arms reasonably well, but here
  we simply show the best results to our molecular ring fitting technique. 
The emission from the second spiral arm (green) extends a little
outside the scope of the observed CO emission. With a lower rotation
curve ($v=220$ km s$^{-1}$), this is avoided. Otherwise there is little
difference for a lower rotation curve.

It is difficult to constrain the starting radius of the
  Sct-Cen Arm, given the complex emission towards the Galactic
  Centre. We can however see that the arm marked as the Perseus arm
  cannot extend much further inward, else there would be unobserved
  emission (at $l\sim15^{\circ}$, $v\sim150$ km s$^{-1}$) for this
  model. 

In the centre panels of Figure~1 we show a 4 armed spiral model. 
To the two armed model, we added a third arm (cyan) using our fitting
technique. This arm naturally reproduces the Outer arm
emission. We tried adding a 4th arm, however our fitting technique did not show any minimum in the expected vicinity.
Since this arm is close
to the Sun (or observer), the emission, and therefore results became
dominated by this arm, whereas for the observations, the strongest
emission coincides with the molecular ring. Therefore we reduced our
calculated emission from this arm by a factor of 10 compared to the
other arms to fit the fourth arm. This arm then reproduces the Carina
arm, and also the tangent point at $l\sim 50^{\circ}$, $v\sim50$ km s$^{-1}$.   
The factor of 10 is somewhat arbitrary but does seem to indicate that
any arm between the Sun and the Sct-Cen arm is somewhat weaker.
Overall though, the addition of extra arms superimposed on the Sct-Cen arm 
contribute further to the emission of the molecular ring.

In the lower panels of Figure~1 we show the results for 
rings. We show 3 examples with radii of 4, 5 and
6 kpc. It can be seen from Figure~1 that a ring does not cover as much
of the emission of the observed `molecular ring' as a spiral arm. The
reason for this is because the observed molecular ring is actually
curved in $v-l$ space, whereas the emission from an
actual ring is a straight line. So we see that whilst a 4 or 5 kpc
radius ring can fit emission for positive longitudes, it misses the
emission at negative longitudes. Conversely, a 6 kpc ring reproduces
the observed emission at negative longitudes, but does not agree with
the brightest observed emission at positive longitudes.
 From our
fitting technique, the 6 kpc ring gave the best fit, although by eye 5
kpc appears best (the difference is probably because the observed
emission extends further at negative longitudes compared to positive
longitudes). 

In Table~1 we show the difference between the model estimated CO
emission and that of the observations,
$\sigma^2=(I_m-I_{obs})^2$ for 1 (corresponding to the Sct-Cen
arm in Figure~1), 
2 (upper panels, Figure~1) and 4 (middle panels, Figure~1) armed spiral models
with a pitch angle of 11$^{\circ}$ and a ring of 6 kpc radius. 
The spiral
models provide better fits statistically compared to the ring model.
The 2 and 4 armed spirals are also better fits compared to the
1 armed model or ring, which is not surprising because they allow more complexity and 
cover a larger area where the observed emission lies. However distinguishing between the 
2 and 4 armed models (and allowing for the different degrees of freedom) is probably beyond the scope of our approach.

We also performed a simpler test to compare between our models and the observed CO emission 
only between the tangent points of the Sct-Cen arm (the `arm' fit). This test corrects for any bias due to 
spiral arms crossing the region $l=\pm 50^{\circ}$ multiple times. We show the results in Table
2, and again the spiral arm models still provide better fits compared to the ring.

\begin{table}
\centering
\begin{tabular}{c|c}
 \hline 
Model & $\sigma^2/\sigma_{4arm}^2$ \\
 \hline
Ring & 2.27 \\
1 armed spiral & 2.02 \\
2 armed spiral & 1.33 \\
4 armed spiral & 1.0 \\
\hline
\end{tabular}
\caption{The difference between the observed and analytic emission,
  $\sigma^2$, where $\sigma^2=(I_m-I_{obs})^2$ is shown for the ring
  model, and spiral arm models with a pitch angle of $11^{\circ}$. The 1
  armed model refers to just the Sct-Cen arm in Figure~1. The
results are normalised to our best fit model, the 4 armed spiral. The
4 armed spiral provides the best fit, though all cases are a better
fit compared to a ring.}
\label{runs}
\end{table}

\begin{table}
\centering
\begin{tabular}{c|c|c}
 \hline 
Pitch angle ($^{\circ}$) & $\sigma^2/\sigma_{4arm}^2$ (`total') &$\sigma^2/\sigma_{4arm}^2$ (`arm') \\
 \hline
8.5 & 2.00 & 2.13 \\
11 & 2.02 & 2.08 \\
13.5 & 2.11 & 2.06 \\
16 & 2.18 & 2.01 \\
\hline
\end{tabular}
\caption{We show the  difference between the observed and analytic emission,
  $\sigma^2$, where $\sigma^2=(I_m-I_{obs})^2$ for the Sct-Cen arm with
  different pitch angles. For the left column, we consider emission
  along the total length of the arm. For the right column, we consider
  emission only between the tangent points
  of the Sct-Cen arm. The results are normalised as for Table~1. The
  best fit pitch angle is dependent on the fitting technique, however
  in all cases, the fit is better compared to a ring (Table~1).}
\label{runs}
\end{table}

\subsection{Parameter study}
In this section, we investigate how altering the parameters of our
models affects how well the spiral arms fit the observed emission. 

The main parameter which we can vary is the pitch angle of the spiral
arms. In Figure~2 we show models with pitch angles of 8.5$^{\circ}$,
13.5$^{\circ}$ and 16$^{\circ}$. From Figure~2 we see that in all cases the
molecular ring is well reproduced. Again, these are our best fit models,
where we have fitted for the orientation of the arms. Thus to a large
extent there is a degeneracy between the pitch angle and the orientation
of the arms, though it is not always possible to reproduce other Galactic features.
\begin{figure*}
\centerline{\includegraphics[scale=0.48, bb=-150 0 530 410]{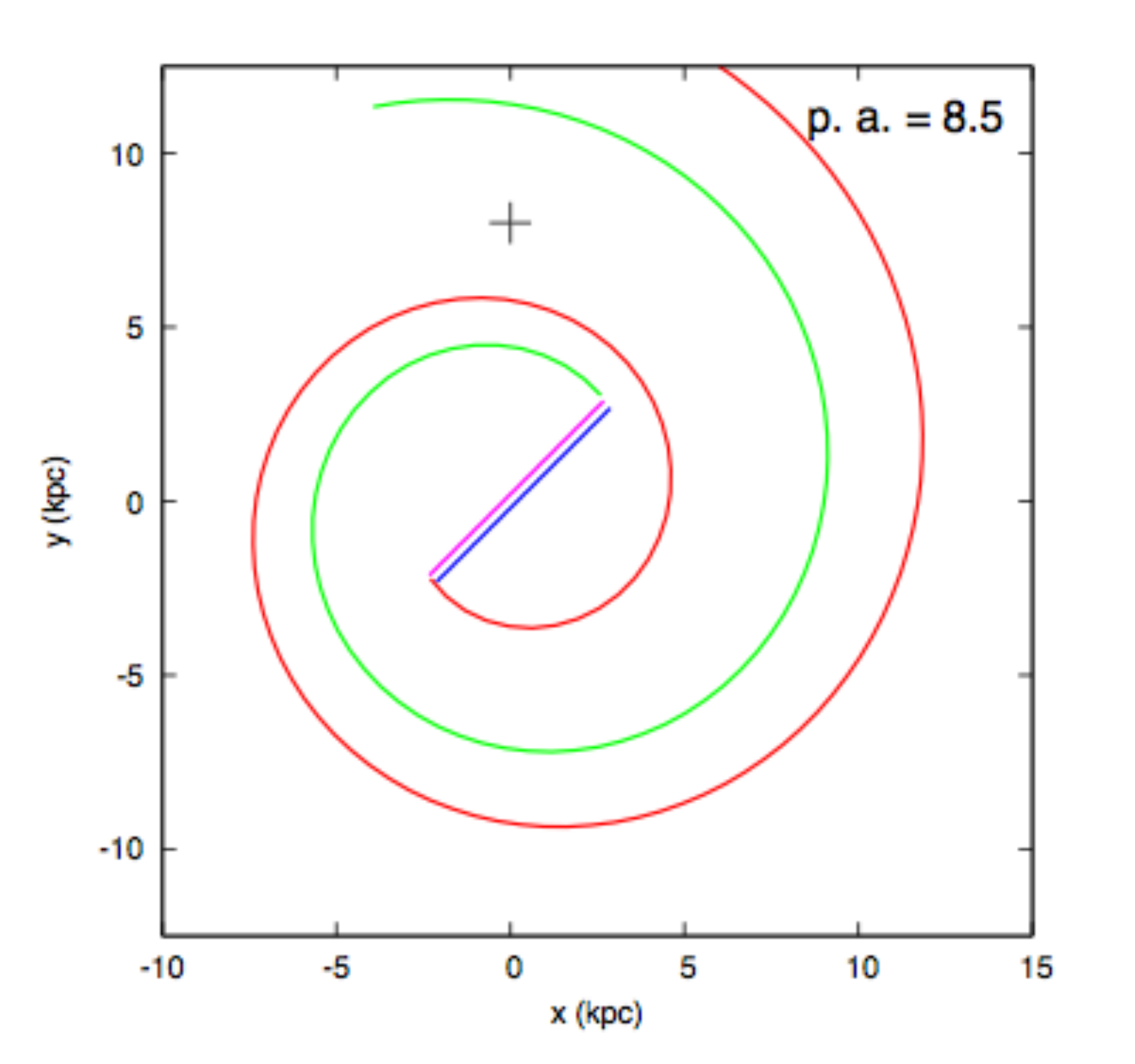}
\includegraphics[scale=0.58, bb=100 0 530 260]{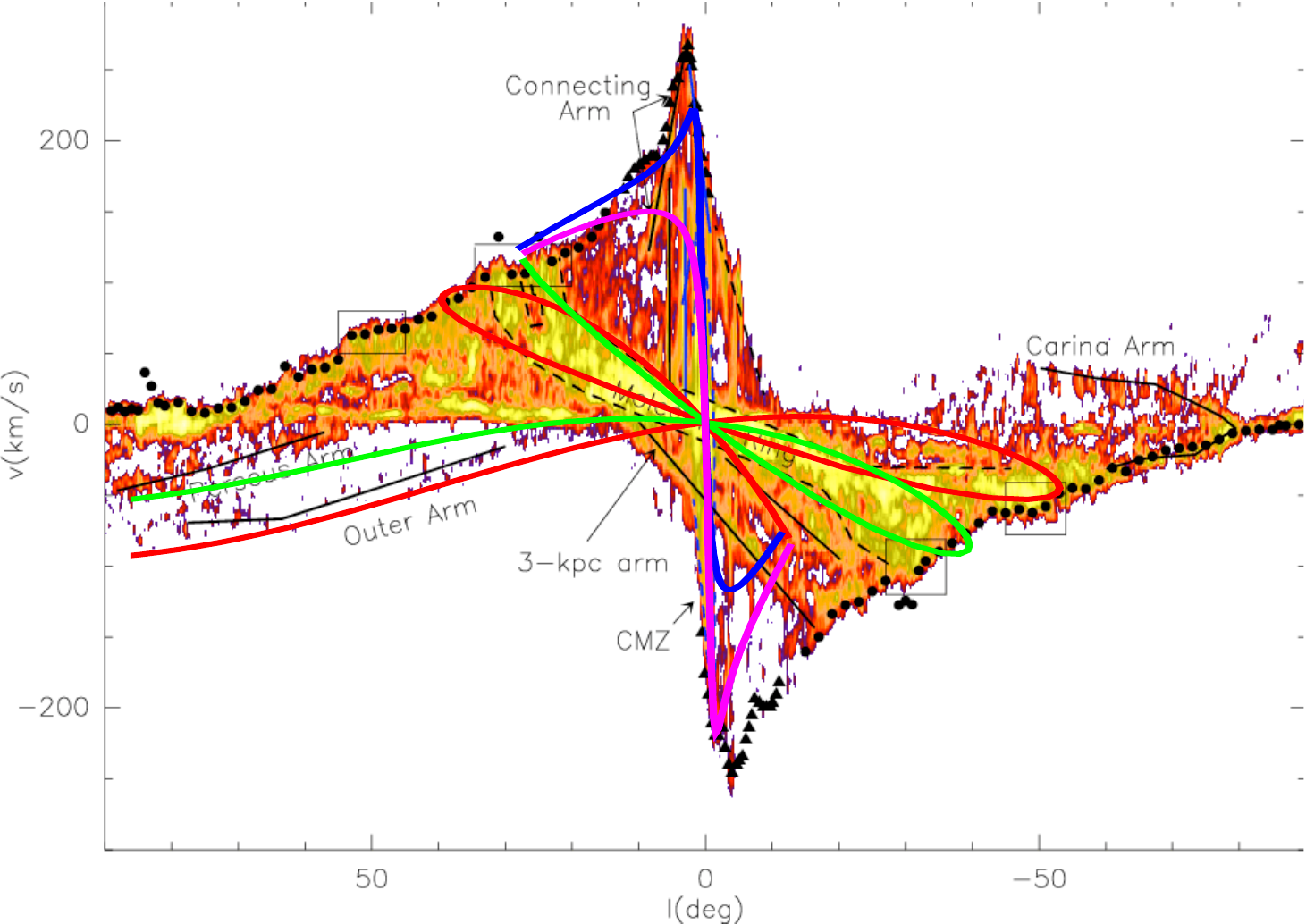}}
\centerline{\includegraphics[scale=0.48, bb=-150 -50 530 380]{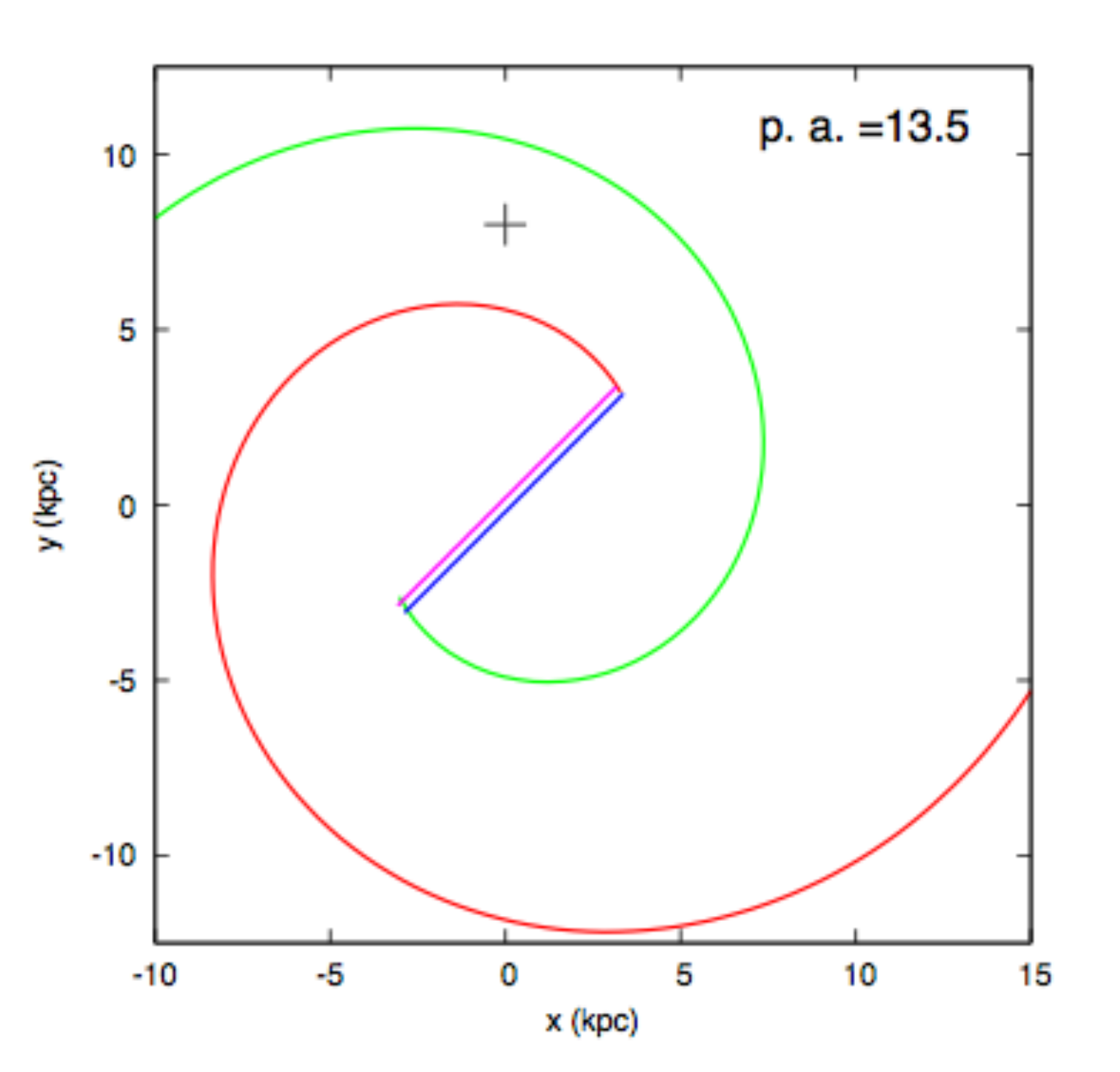}
\includegraphics[scale=0.58, bb=100 -35 530 280]{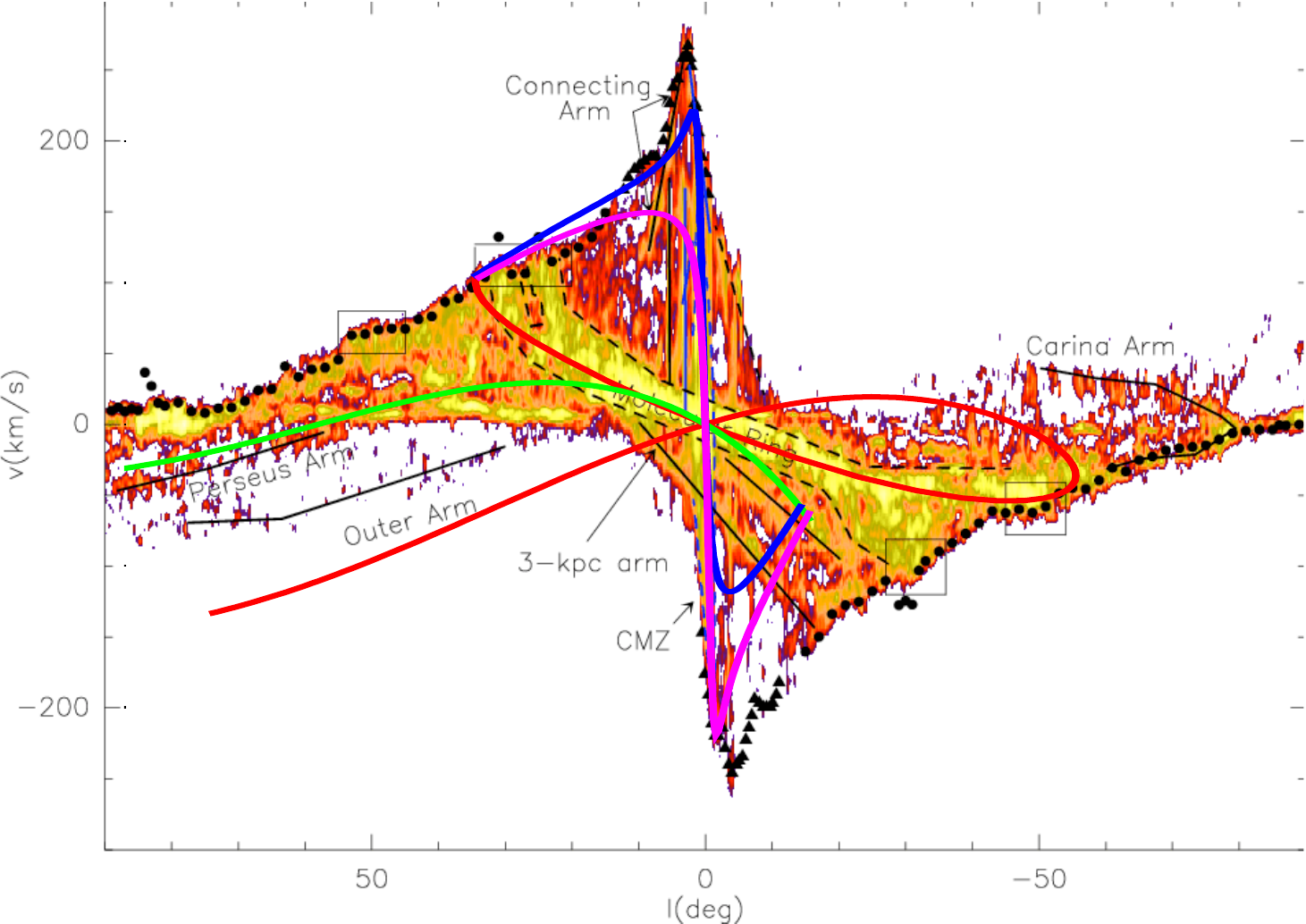}}
\centerline{\includegraphics[scale=0.48, bb=-150 -50 530 330]{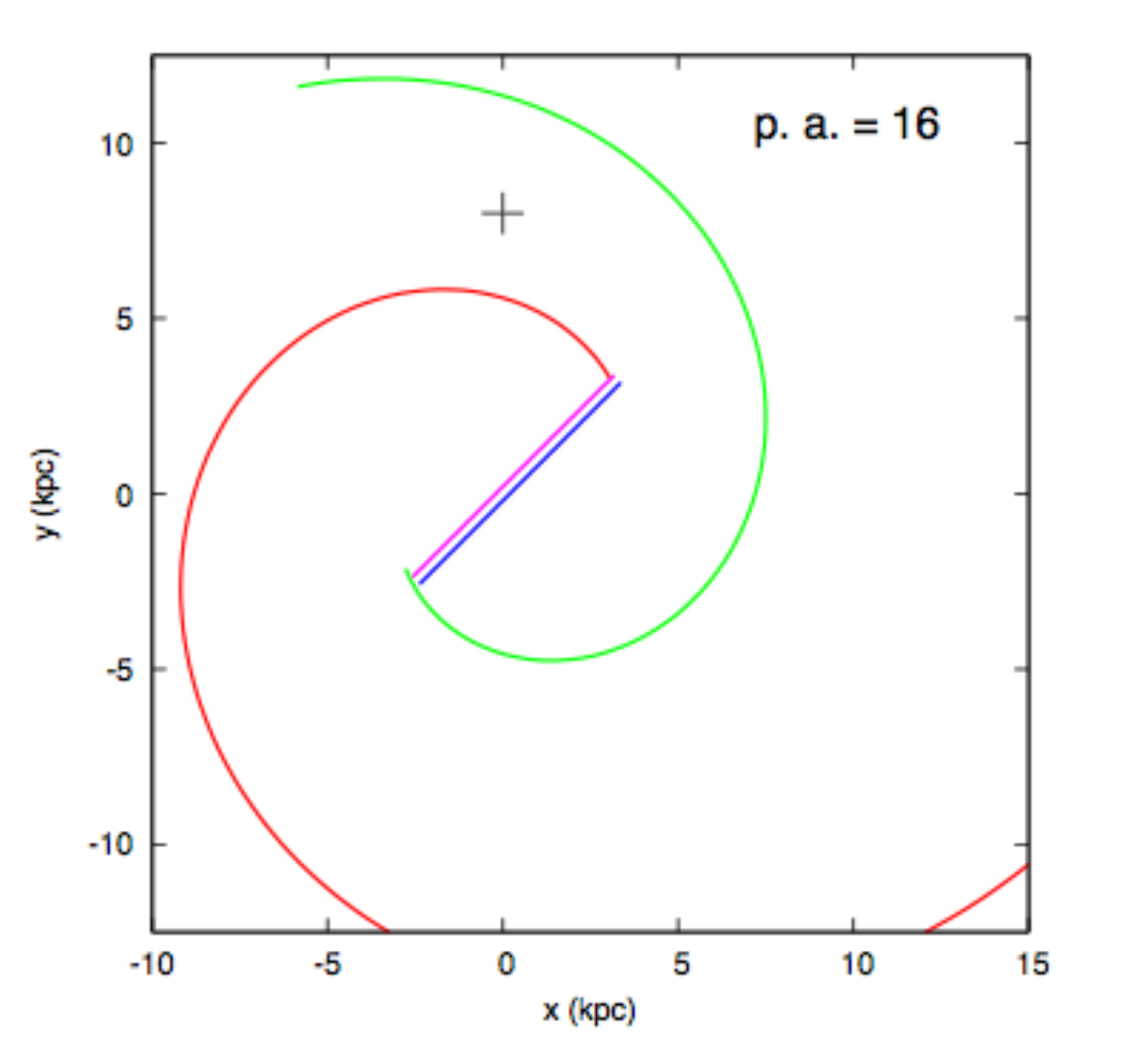}
\includegraphics[scale=0.58, bb=100 -35 530 230]{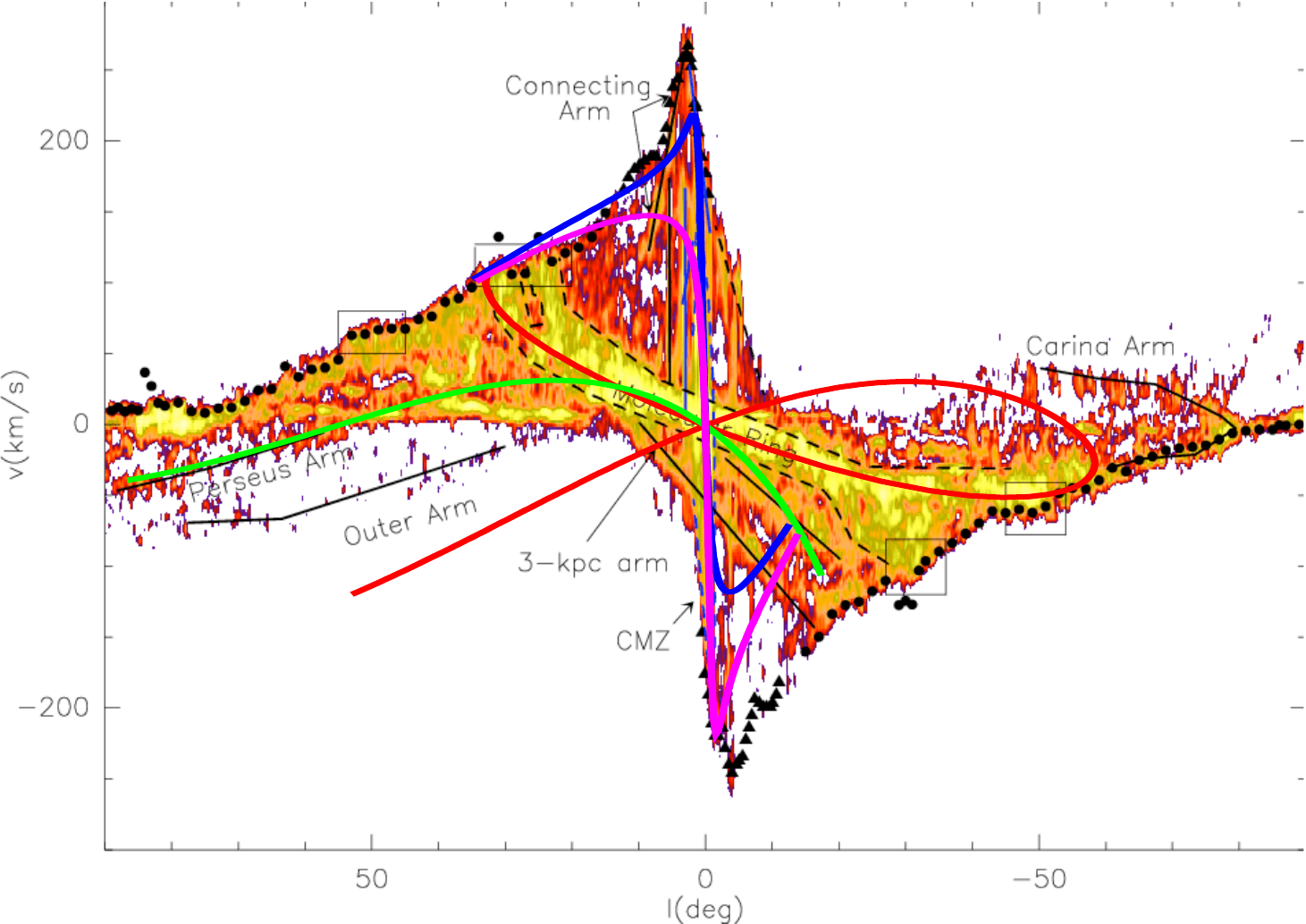}}
\caption{The spiral pattern is shown for models with 2 spiral
  arms and pitch angles of 8.5$^{\circ}$ (top), 13.5$^{\circ}$ (centre) and 16$^{\circ}$
  (lower). The position of the observer is marked by the cross at 8 kpc.
The arms have
  been rotated to provide a best fit to the data. In all cases the
  Sct-Cen arm easily fits the molecular ring. For a pitch angle of
  $8.5^{\circ}$, there is a good agreement with the observed CO emission,
  but the spiral arms are highly asymmetric. For higher pitch angles,
 we do not reproduce
  the tangent point at $l=30^{\circ}$, $v=-100$ km s$^{-1}$ as
  the arms do not extend to the inner part of the Galaxy, and also  
it is difficult to reproduce this tangent point and the
  Perseus arm simultaneously. Please view journal online for colour versions of the figures.}
\end{figure*}

For the pitch angle of $8.5^{\circ}$, we obtain a very asymmetric
pattern. Our fitting technique does not take into
account tangent points, or the Perseus arm which are also constraints
on the second spiral arm (green). However even if we adjust the second
spiral to fit the tangent point at $l\sim-30^{\circ}$, $v\sim-100$ km
s$^{-1}$, 
the structure is still highly
asymmetric. 

For pitch angles of $13.5^{\circ}$ and $16^{\circ}$ we do not produce the tangent
point at $l\sim-30^{\circ}$, $v\sim-100$ km
s$^{-1}$ simply because the emission does not extend that far
inwards. We could continue the arms for another half rotation, however
this would lead to a rather short ($\lesssim 2$ kpc) bar, with the
assumption that the
arms begin at the bar. In any case
it is difficult to match the tangent point and the Perseus arm
simultaneously, especially when the pitch angle is $16^{\circ}$.
Another constraint is the observed distance to the
Perseus arm, which is around 2 kpc towards $l=134^{\circ}$ \citep{Xu2006}. We get
good agreement with this for the models we present.
However if we
constrain the Perseus spiral arm to match the the tangent point at $l\sim-30^{\circ}$, $v\sim-100$ km
s$^{-1}$, the distance to the Perseus arm becomes too large for the
higher pitch angles. 
 Finally we note that we do not get good agreement with the newly
observed outer HI arm \citep{Dame2011} with these pitch angles, in
particular $8.5^{\circ}$ and $16^{\circ}$. 

We show in Table~2 how well the models with different pitch angles fit
the observed data, using the `total' and `arm' fits. 
The `total' fit favours lower pitch angles, because the arms cover
more of the region of observed emission. The `arm' fit favours larger
pitch angles, and provides a best fit pitch angle of $16.5^{\circ}$. Both
techniques however neglect features such as the Perseus arm, outer HI
arm, and supposed tangent points, which appear necessary 
to constrain the pitch angle.  With these extra constraints,
the $11^{\circ}$ pitch angle reproduces more of the CO features.

We also tested the orientation of the bar, the rotation curve and the position of the
observer in the Galaxy. As noted before, decreasing the rotation curve
to 220 km s$^{-1}$ slightly reduces the scope of the emission in
velocity space. Moving the radius of the
observer closer to the Galactic Centre increases the span of the
nearer spiral arm in the velocity plot. Thus to still achieve a
similar pattern to Figure~1, a lower pitch angle would be required,
and likewise if the observer is further out in the Galaxy, a higher
pitch angle would be needed. So again there is a degeneracy
between the Galocentric radius and the pitch angle, but again,
with significant departures from the observed values, it is more difficult to
reproduce all the features in the CO map simultaneously.

Changing the orientation of the bar does
not make a very noticeable difference to the location of emission
associated with the bar. It is difficult to reproduce the full extent
of CO emission in velocity space associated with the bar, i.e.\ for
velocities in excess of $+200$ km s$^{-1}$. This could be due to a higher
rotation curve, motions along the bar, or simply features near the
Galactic Centre that we are missing.

\subsection{Comparison to other models of the Milky Way}
There have been several models suggested for the structure of the
Milky Way in recent years. \citet{Vallee2005} proposes a 4 armed spiral model with a
pitch angle of 12$^{\circ}$. The main difference between his model, and
our 4-armed model is that we suppose that the Perseus and
Scutum-Centaurus Arm start at the bar, whereas \citet{Vallee2005}
assumes that the Sagittarius-Carina and Cygnus (Outer) Arms start at
the bar (these are the equivalent to the arms coloured cyan and yellow
on Figure~1). We constrained the models such that the Scutum-Centaurus
and Perseus Arm begin at the bar, because these arms are seen in the
stars and the gas, whereas the other arms may not be associated with
stellar enhancements \citep{Drimmel2000,Benjamin2008}.

\citet{Churchwell2009} also
proposed a schematic of the Galaxy based on the GLIMPSE infrared
survey (see also \citealt{Benjamin2008}). They propose that the Galaxy is a 2-armed spiral, the main arms
being the Scutum-Centaurus and Perseus Arms, with several secondary
spiral arms. They
adopt a long bar, but do not state the pitch
angle of the arms. This is similar to our 11$^{\circ}$ pitch angle model
if we choose a longer bar, and neglect the first 180$^{\circ}$ rotation of the
spiral arms, or the models we show for larger pitch angles. We did note
though in Section 3.1 that starting the arms further out in the disc
would likely miss regions of emission in the CO $l-v$ diagram.  

Finally, \citet{Steiman2010} propose a model based on [C\textsc{II}] and [N\textsc{II}]
cooling lines. This is very similar to our 4 armed model in
Figure~1. The main difference is that they use slightly larger pitch
angles ($13-16^{\circ}$), and whilst we obtained a reasonable fit with one
pitch angle, they used different pitch angles for each arm. With the
pitch angle for the Scutum-Centaurus arm they used 
($15.5^{\circ}$), it is difficult to reproduce the HI feature seen by
\citet{Dame2011} as the arm barely extends to the third Galactic quadrant. 

\section{Discussion}
It is relatively easy to find a spiral arm configuration such that the
region corresponding to the molecular ring is reproduced by a nearby
spiral arm. If we rotated the nearest spiral arm (with respect to the
Galactic Centre), we would obtain a similar feature with a different
gradient. Increasing or decreasing the pitch angle of the arms changes
the extent of the molecular ring. Other spiral arms enhance the
molecular ring, as all overlap at least at $l=0$. Finally, as the
molecular gas surface density decreases with radius, outside the
vicinity of the bar, the maximum emission will be from the near spiral
arm close to the bar. 

Whilst a ring feature can also represent CO emission similar to the
observed molecular ring, a ring does not fit the observations as well
as a spiral arm. This is because the spiral arm appears curved in
$v-l$ space, similar to the observed molecular ring, but dissimilar to
a true ring. Thus whilst we cannot rule out that the molecular ring
corresponds to a true ring, we found that a spiral arm produced a
better fit compared to a ring over all our range of pitch angles, and
this finding was robust to the details of our fitting technique.

\citet{Binney1991} proposed that the molecular ring 
could be due to the outer Lindblad resonance of the bar.
We also
performed simulations with a barred potential to examine whether a
gaseous ring would form. However generally gas features produced at
the end of the bar 
are highly elliptical, as seen already in numerical simulations
(e.g.\ \citet{Wada2001,Lin2008}). The elongated features due to the bar
in our Galaxy
may well correspond to the far and near 3 kpc arms, seen in the
molecular gas data. In our simulations of bars,
any feature corresponding to a ring was again simply the spiral
arms close to the Galactic centre. In fact, few galaxies show obvious
rings in the gas at the end of the bar -- typically
rings are nuclear rings much nearer to the centre, or features caused
by large collisions. Our Galaxy is probably not unusual in this respect.

Our fiducial model adopts a pitch angle of $11^{\circ}$, assuming a
Galocentric distance of 8 kpc. All the pitch angles we tried
could reproduce a feature similar to the molecular ring, and from fitting
the molecular ring alone we find a best fit pitch angle of 16.5$^{\circ}$. However deviations from the $11^{\circ}$ model tend to
reproduce fewer of the other observed features in the CO $l-v$ diagram,
or produce highly asymmetric arms. Larger pitch angles seem to point
towards a longer bar, which means CO in the lower right quadrant of
the $v-l$ plot is absent. Moreover the supposed tangent point of the
inner part of the Perseus arm, and the distance to the Perseus arm,
cannot be matched simultaneously with large pitch
angles.
Our 2 armed spiral model is fairly symmetric, and is thus
consistent with a density wave originating at the bar. Though we cannot 
rule out that the Galaxy simply consists of several, asymmetric, 
spiral arms, it seems less likely that by coincidence they match both the inner and outer spiral
structure simultaneously.
 Reproducing all the features in the CO,
e.g. Outer Arm, Carina Arm, requires 4 spiral
arms. 

It is thought that the Galaxy may exhibit two spiral arms which are evident in both stars and gas, whilst
other features, e.g.\ the Sagittarius Arm, are only seen in the gas
\citep{Drimmel2000,Benjamin2008}. 
The next step would be to try and
produce hydrodynamical models with potentials based on the 2 armed
spiral pattern shown here for example, and see whether there are
gaseous spurs or arms which do not correspond to stellar features
and whether they correspond to observed
features in CO\@. Such features could arise from the shearing of clouds
in the spiral arms \citep{Kim2002,Dobbs2006} or resonances in the disc
\citep{Patsis1997,Chak2003,Martos2004}.

\section{Acknowledgments}
We thank the referee Tyler Foster for suggestions that have
substantially improved the paper.
We are very grateful to Nemesio Rodriguez for providing the CO
velocity longitude plot from \citet{Rod2008}, and to Tom Dame for
providing the CO velocity-longitude data used for our comparison
tests. We also thank Panos Patsis for valuable discussions.
CLD acknowledges funding from the European Research Council for the 
FP7 ERC starting grant project LOCALSTAR.

\bibliographystyle{mn2e}
\bibliography{Dobbs}

\bsp
\label{lastpage}
\end{document}